\documentstyle[11pt]{article}
\begin{document}
\title{Modeling the viscoelastoplastic behavior of amorphous
glassy polymers}
\author{Aleksey D. Drozdov\\
Institute for Industrial Mathematics\\
4 Hanachtom Street\\
Beersheba, 84311 Israel}
\date{}
\maketitle

\begin{abstract}
Constitutive equations are derived for the viscoelastoplastic
response of amorphous glassy polymers at isothermal loading
with small strains.
A polymer is treated as an ensemble of cooperatively relaxing regions (CRR)
which rearrange at random times as they are thermally agitated.
Rearrangement of CRRs reflects the viscoelastic response of the bulk medium.
At low stresses, CRRs are connected with each other,
which implies that the macro-strain in a specimen coincides
with micro-strains in individual relaxing regions.
When the average stress exceeds some threshold level,
links between CRRs break and relaxing domains begin to slide
one with respect to another.
Sliding of micro-domains is associated with the viscoplastic behavior
of polymers.
Kinetic equations are proposed for viscoplastic strains
and for the evolution of the threshold stress.
These equations are validated by comparison with experimental data
in tensile relaxation tests and in tests with constant strain rates.
Fair agreement is demonstrated between results of
numerical simulation and observations for a polyurethane resin
and poly(methyl methacrylate).
\end{abstract}


\section{Introduction}

A constitutive model is derived for the viscoelastoplastic
response of amorphous glassy polymers at isothermal loading
with small strains.
The time-dependent behavior of glassy polymers has become the focus
of attention in the past decade \cite{PB91}--\cite{ZK00}.
This may be explained by the use of new experimental techniques
which allow complicated time-dependent programs of loading
to be carried out (loading and unloading with
different strain rates interrupted by relaxation and
superposition of small oscillatory strains on loading
with constant strain rates) at various temperatures.

The present study is concerned with the response of polymers
in relaxation tests and in tests with constant strain rates.
We confine ourselves to relatively slow programs of loading
and to temperatures in the sub--$T_{\rm g}$ region, where $T_{\rm g}$
is the glass transition temperature.
This implies that crazing and other phenomena associated
with brittle fracture are excluded from our consideration,
which focuses on the sub-yield and post-yield response of ductile polymers.
To simplify the analysis, we deal with uniaxial loading at small strains.
These limitations may be avoided by using conventional approaches,
see, e.g., \cite{Dro96}, but we do not dwell on this issue.
The purpose is to develop a constitutive model which describes
the viscoelastic and viscoplastic response of amorphous polymers
at the micro-level, on the one hand, and which is sufficiently simple
to be used in the study of three-dimensional deformations
of polymers with complicated geometry, on the other.
We aim to design stress--strain relations which correctly predict
two phenomena observed in experiments:
(i) changes in elastic moduli and relaxation spectra induced
by changes in the test's temperature
and (ii) the influence of the strain rate and temperature
on the apparent yield stress.

To describe the viscoelastic response of glassy polymers, we employ
the concept of cooperative relaxation \cite{AG65}.
An amorphous polymer is treated as an ensemble of cooperatively
rearranging regions (CRR).
Any CRR is modeled as a globule consisting of scores of neighboring
strands of long chains which change their position
simultaneously because of large-angle reorientation of
strands \cite{Dyr95,Sol98}.
The characteristic length of a relaxing region amounts to several
nanometers \cite{RN99}.
In the phase space, a CRR is treated as a point trapped in its
potential well on the energy landscape.
A CRR spends most time at the bottom level of its trap.
At random times, it hops to higher energy levels being thermally
agitated.
We adopt the transition-state theory \cite{Gol69} which postulates
that some liquid-like (reference) level exists on the energy landscape.
When a CRR reaches this reference state in a hop,
it totally relaxes.
Viscoelasticity of a glassy polymer is modeled as sequential
rearrangements of CRRs trapped in potential wells with various depths.

CRRs are assumed to be connected with one another by some links
(which model temporary crosslinks, entanglements and van der Waals
forces).
At relatively low stresses, the links prevent mutual displacements of
CRRs, which implies that the micro-strains in relaxing regions coincide
with the macro-strain in a specimen.
When the average stress in an ensemble of CRRs exceeds some threshold
value, links between CRRs break, and relaxing regions become
(partially) free to change their positions with respect to each other.
The mutual displacement of CRRs models the viscoplastic deformation at
the micro-level.

The exposition is organized as follows.
Thermally induced rearrangement of CRRs is discussed in Section 2.
Section 3 deals with constitutive relations for the viscoelastic
behavior of an amorphous polymer.
Kinetic equations for the evolution of the threshold stress
are proposed in Section 4.
Section 5 is concerned with comparison between experimental data and
results of numerical simulation.
Some concluding remarks are formulated in Section 6.

\section{Kinetics of rearrangement}

The position of the liquid-like state at the glass transition
temperature $T_{\rm g}$ is accepted as the zero energy level
on the energy landscape.
The depth of a potential well where a CRR is trapped
with respect to the zero energy level is denoted as $w\geq 0$.
We suppose that the position of the reference state on
the energy landscape may change with temperature
and denote by $\Omega(T)$ its ascent at some temperature $T<T_{\rm g}$
with respect to the position of the reference level at $T_{\rm g}$
(the function $\Omega(T)$ satisfies the natural
condition $\Omega(T_{\rm g})=0$).

Denote by $X_{0}$ the concentration of traps per unit mass
of the bulk medium.
For simplicity, $X_{0}$ is assumed to be a constant
(independent of temperature and the loading history).
Rearrangement of relaxing regions is entirely characterized
by the function $X(t,\tau,w)$ which equals the concentration of CRRs
(per unit mass at time $t$)
trapped in cages with potential energy $w$ which
have last rearranged before instant $\tau\leq t$.
The current distribution of traps is described by the
probability density, $p(t,w)$, of traps with energy $w$
at time $t$.
The functions $p(t,w)$ and $X(t,\tau,w)$ are connected by the formula
\begin{equation}
p(t,w)=\frac{X(t,t,w)}{X_{0}}.
\end{equation}
In the sequel, we confine ourselves to equilibrated polymers
with steady-state distribution of traps.
This means that the function $p$ is independent of time.
Bearing in mind that in the sub--$T_{\rm g}$ region,
the distribution of traps with various energies
drastically depends on temperature $T$, we set $p=p(T,w)$.

Let $q(\omega)d\omega$ be the probability for a CRR
to reach (in a hop) the energy level that exceeds the bottom level
of its cage by some value $\omega^{\prime}$ located in the
interval $[ \omega,\omega+d\omega ]$.
Referring to \cite{BM97}, we set
\[ q(\omega)=\alpha\exp(-\alpha\omega), \]
where $\alpha$ is a material constant.
The probability to reach the reference state in an arbitrary hop
for a polymer equilibrated at a temperature $T$ reads
\[ Q(T,w) =\int_{w+\Omega(T)}^{\infty} q(\omega)d\omega
=\exp \biggl [-\alpha \Bigl (w+\Omega(T) \Bigr )\biggr] . \]
Denote by $\Gamma_{0}=\Gamma_{0}(T)$ the attempt rate
(the average number of hops in a trap per unit time).
The rate of rearrangement $R$ equals the product of the attempt rate
$\Gamma_{0}$ by the probability of reaching the reference state in a hop $Q$,
\begin{equation}
R(T,w)=\Gamma(T) \exp (-\alpha w),
\end{equation}
where the relaxation rate $\Gamma(T)$ is given by
\begin{equation}
\Gamma(T)=\Gamma_{0}(T) \exp \Bigl [ -\alpha \Omega(T) \Bigr ].
\end{equation}
Equating the relative rates of rearrangement to the function $R$,
we arrive at the differential equations
\begin{equation}
\frac{1}{X(t,0,w)} \frac{\partial X}{\partial t}(t,0,w) = -R(T,w),
\qquad
\biggl [ \frac{\partial X}{\partial \tau} (t,\tau,w)\biggr ]^{-1}
\frac{\partial^{2} X}{\partial t\partial \tau}(t,\tau,w)
= -R(T,w).
\end{equation}
The solutions of Eq. (4) with the initial condition
$X(0,0,w)=X_{0} p(T,w)$ [see Eq. (1)] read
\begin{eqnarray}
X(t,0,w) &=& X_{0}p(T,w) \exp \Bigl [ -R(T,w)t \Bigr ],
\nonumber\\
\frac{\partial X}{\partial \tau}(t,\tau,w)
&=& X_{0} F(\tau,w) \exp \Bigl [ - R(T,w)(t-\tau) \Bigr ],
\end{eqnarray}
where
\[ F(\tau,w)=\frac{1}{X_{0}}
\frac{\partial X}{\partial \tau}(t,\tau,w)\biggl |_{t=\tau}. \]
The number $N(t,w)$ of relaxing regions (per unit mass) in traps with
potential energy $w$ rearranged within the interval $[t,t+dt ]$
equals the sum of the numbers of CRRs rearranged for the first time
at instant $t$ and those that have last rearranged
at some instant $\tau\in [0,t)$ and reach the reference state at time $t$,
\begin{equation}
N(t,w)=-\frac{\partial X}{\partial t}(t,0,w)
-\int_{0}^{t}\frac{\partial^{2} X}{\partial t\partial \tau}
(t,\tau,w) d\tau .
\end{equation}
Neglecting the duration of a hop (a few picoseconds \cite{Dyr95})
compared to the characteristic time of relaxation in the
sub--$T_{\rm g}$ region, we find that $N(t,w)$ coincides with
the number of CRRs landing within the interval $[t,t+dt ]$
in traps with potential energy $w$,
\begin{equation}
N(t,w)=X_{0}F(t,w).
\end{equation}
Substitution of Eqs. (4) to (6) into Eq. (7) results in the Volterra
equation
\begin{equation}
F(t,w) = R(T,w)\biggl \{ p(T,w)\exp\Bigl [ -R(T,w)t \Bigr ]
+\int_{0}^{t} F(\tau,w) \exp \Bigl [ - R(T,w)(t-\tau) \Bigr ]
d\tau\biggr \}.
\end{equation}
The solution of Eq. (8) is given by
\[ F(t,w)=R(T,w)p(T,w), \]
which, together with Eq. (5) implies that
\begin{equation}
\frac{\partial X}{\partial \tau}(t,\tau,w)
= X_{0} R(T,w)p(T,w) \exp \Bigl [ - R(T,w)(t-\tau) \Bigr ].
\end{equation}

\section{Constitutive equations}

We assume that the macro-strain $\epsilon$ equals the sum
of the viscoelastic strain $\epsilon_{\rm e}$ and the viscoplastic
strain $\epsilon_{\rm p}$,
\begin{equation}
\epsilon(t)=\epsilon_{\rm e}(t)+\epsilon_{\rm p}(t).
\end{equation}
The viscoelastic strain, $\epsilon_{\rm e}$, is the average strain
in relaxing regions, and the viscoplastic strain, $\epsilon_{\rm p}$,
is the average strain induced by displacement of CRRs one with respect
to another.
To simplify our analysis, we neglect the distribution of micro-strains
and suppose that the micro-strain in any CRR coincides with
$\epsilon_{\rm e}$.

A CRR is assumed to totally relax when it reaches
the liquid-like state, which means that the natural (stress-free)
configuration of a CRR after rearrangement coincides with
the actual configuration of the bulk medium at the instant of rearrangement.
For uniaxial deformation, the strain from the natural configuration
of a relaxing region to its actual configuration at time $t$ is
given by
\begin{equation}
\varepsilon(t,\tau)=\epsilon_{\rm e}(t)-\epsilon_{\rm e}(\tau),
\end{equation}
where $\tau\leq t$ is the last instant when the region was rearranged.
A CRR is modeled as a linear elastic medium with the mechanical energy
\[ \phi (t,\tau)=\frac{1}{2} c \varepsilon^{2}(t,\tau), \]
where $c=c(T)$ is the rigidity of a relaxing region.
To minimize the number of material parameters in the constitutive equations,
we assume the rigidity $c$ to be constant and independent of the energy
of trap $w$.
Summing the mechanical energies of CRRs and neglecting the energy
of their interaction,
we find the strain energy density of a polymer (per unit mass)
\[ \Phi(t) = \frac{1}{2}c \biggl [ \varepsilon^{2} (t,0)
\int_{0}^{\infty} X(t,0,w) dw
+\int_{0}^{t} \varepsilon^{2}(t,\tau)d\tau
\int_{0}^{\infty} \frac{\partial X}{\partial \tau}(t,\tau,w) dw\biggr ]. \]
Substitution of expressions (10) and (11) into this equality results in
the formula
\begin{eqnarray}
\Phi(t) &=& \frac{1}{2}c \biggl [ \Bigl (\epsilon(t)
-\epsilon_{\rm p}(t)\Bigr )^{2} \int_{0}^{\infty} X(t,0,w) dw
\nonumber\\
&& +\int_{0}^{t} \Bigl ( (\epsilon(t)-\epsilon(\tau))-
(\epsilon_{\rm p}(t)-\epsilon_{\rm p}(\tau))\Bigr )^{2} d\tau
\int_{0}^{\infty} \frac{\partial X}{\partial \tau}(t,\tau,w) dw\biggr ].
\end{eqnarray}
At small strains, the average stress $\sigma(t)$ is expressed in terms
of the macro-strain $\epsilon(t)$ by the conventional formula
\begin{equation}
\sigma(t)=\rho \frac{\partial \Phi(t)}{\partial \epsilon(t)},
\end{equation}
where $\rho$ is mass density in the stress-free state.
Combining Eqs. (12) and (13), we arrive at the stress--strain relation
\begin{eqnarray}
\sigma(t) &=& E_{0} (T) \biggl [ \Bigl (\epsilon(t)
-\epsilon_{\rm p}(t)\Bigr )\int_{0}^{\infty} X(t,0,w) dw
\nonumber\\
&& +\int_{0}^{t} \Bigl ( (\epsilon(t)-\epsilon(\tau))-
(\epsilon_{\rm p}(t)-\epsilon_{\rm p}(\tau))\Bigr ) d\tau
\int_{0}^{\infty} \frac{\partial X}{\partial \tau}(t,\tau,w) dw\biggr ],
\end{eqnarray}
where $E_{0}(T)=\rho c(T)X_{0}$ is the initial Young modulus.

To proceed with the analysis of Eq. (14), a concrete form of the distribution
function $p(T,w)$ should be established.
Referring to the random energy model \cite{RB90},
we assume that the probability density $p$ is quasi-Gaussian,
\begin{equation}
p(T,w)=C(T)\exp \biggl [-\frac{(w-W(T))^{2}}{2\Sigma^{2}(T)}\biggr ]
\quad (w\geq 0),
\qquad
P(T,w)=0 \quad (w<0),
\end{equation}
where $W(T)$ and $\Sigma(T)$ are parameters of the Gaussian distribution
and the constant $C$ is found from the condition
\[ \int_{0}^{\infty} p(T,w) dw=1. \]

\section{Kinetics of plastic flow}

To describe the evolution of the plastic strain $\epsilon_{\rm p}(t)$,
we propose constitutive equations similar (to some extent) to the
viscoplasticity theory based on overstress \cite{Kre87}
(back stress \cite{BPA88}), see also recent publications
\cite{KB98,KK00}.
For simplicity, we confine ourselves to active loading with
$\dot{\epsilon}(t)=d\epsilon(t)/dt\geq 0$, which means that
the internal parameter $\lambda$ (the material time associated with
loading) may be set equal to the macro-strain $\epsilon$.
Our approach is based on the following hypotheses:
\begin{enumerate}
\item
The level of plastic strain is determined by the threshold stress
$g$ which depends on temperature and the loading history.

\item
Plastic deformation occurs only when the current average stress $\sigma(t)$
exceeds the threshold stress $g(t)$.
The rate of change in the viscoplastic strain $\epsilon_{\rm p}$
(with respect to the internal time $\lambda$)
is proportional to the difference $\sigma-g$,
\begin{equation}
\frac{d\epsilon_{\rm p}}{d\lambda}=K_{1}{\cal H}(\sigma-g),
\end{equation}
where $K_{1}$ is a material parameter and
${\cal H}(x)$ is the Heaviside function,
\[ {\cal H}(x)=\left \{\begin{array}{ll}
1, & x\geq 0,
\\
0, & x<0.
\end{array}\right . \]

\item
For a stress-free material, the threshold stress has a fixed value
$g_{0}$.
At the micro-level, $g_{0}$ determines the strength of links
between CRRs which prevent their mutual displacements.
In the general case, the quantity $g_{0}$ depends on temperature $T$
and on the strain rate $\dot{\epsilon}$ with which a specimen is loaded.
A decrease in $g_{0}$ with temperature is explained
within the framework of the theory of thermally activated processes:
the growth of temperature implies an increase in the amplitude of
thermal fluctuations, which reduces the strength of links between CRRs.
The effect of strain rate is also quite natural provided
that the links break according to the brittle pattern.

\item
When the current stress $\sigma$ exceeds the threshold stress,
the quantity $g$ changes.
The rate of change in the threshold stress is determined by two
mechanisms.
The first is associated with a decrease in the  current
value of $g$ caused by partial destruction of links between CRRs
(we suppose that some distribution of the links' strength exists,
and the links with low strengths break first).
The rate of decrease in the threshold stress (with respect
to the internal time $\lambda$) is assumed to be proportional
to the difference $\sigma-g$.
Bearing in mind that $g$ is a nonnegative function, we set
\begin{equation}
\frac{dg}{d\lambda}=-K_{2}g{\cal H}(\sigma-g),
\end{equation}
where $K_{2}$ is a constant.
Equation (17) ensures that $g(t)=0$ for any $t\geq t_{0}$,
provided that $g(t_{0})=0$.

\item
The other mechanism for changes in the threshold stress $g$ is
associated with thermally induced healing of links between CRRs.
We suppose that this process is similar to craze healing in glassy
polymers, see \cite{PD89} and the references therein.
Because the healing process is driven by thermal fluctuations,
its kinetics is analogous to that for the reformation process
in a temporary network \cite{TE92}.
This means that the rate of reformation (per unit time $t$)
for links between CRRs is proportional to the number of broken links,
\begin{equation}
\frac{dg}{dt}=K_{3}{\cal H}(g_{0}-g),
\end{equation}
where $K_{3}$ is a constant.

\item
In general, the parameters $K_{i}$ ($i=1,2,3$) depend on temperature $T$
and the loading history.
To reduce the number of adjustable parameters in the model, we assume
that the effect of these factors is rather weak and treat $K_{i}$ as
material constants.
\end{enumerate}
When $\lambda$ coincides with the macro-strain $\epsilon$,
Eqs. (16) to (18) are transformed into the nonlinear differential
equations
\begin{eqnarray}
\frac{d\epsilon_{\rm p}}{dt}=K_{1}{\cal H}(\sigma-g)\frac{d\epsilon}{dt},
\qquad
\frac{dg}{dt}=K_{3}{\cal H}(g_{0}-g)-K_{2}g{\cal H}(\sigma-g)
\frac{d\epsilon}{dt}
\end{eqnarray}
with the initial conditions
\begin{equation}
\epsilon_{\rm p}(0)=0,
\qquad
g(0)=g_{0}.
\end{equation}
Given a loading program, $\epsilon(t)$, Eqs. (2), (14), (15), (19) and (20)
uniquely determine the average stress $\sigma(t)$.
The constitutive equations are determined by four functions of
temperature, $E_{0}(T)$, $W(T)$, $\Sigma(T)$ and $\Gamma(T)$,
one function of temperature and strain rate, $g_{0}(T,\dot{\epsilon})$,
and three constants $K_{1}$, $K_{2}$ and $K_{3}$
(without loss of generality, the parameter $\alpha$ in Eq. (2) may be
set to unity).
The total number of adjustable parameters in the model is essentially
less than that in conventional stress--strain relations in
viscoelastoplasticity based on the overstress concept \cite{BK92,KB98,KK00}.

\section{Comparison with experimental data}

To verify the constitutive equations, we determine adjustable parameters
by fitting observations in tensile relaxation tests and in
tensile tests with constant strain rates.
We begin with the standard isothermal relaxation test with
the loading program
\begin{equation}
\epsilon(t)=\left \{ \begin{array}{ll}
0, & t<0,\\
\epsilon_{0}, & t>0,
\end{array}\right .
\end{equation}
where $\epsilon_{0}$ is a given strain (which is assumed not to exceed
the yield strain).
It follows from Eqs. (19) and (20) that
\[ \epsilon_{\rm p}(t)=0,\qquad g(t)=g_{0}(T,0). \]
Substituting these expressions and Eqs. (2), (15) and (21) into Eq. (14),
we find that
\[ E(T,t)=C(T) E_{0}(T) \int_{0}^{\infty} \exp \biggl [
-\biggl ( \frac{(w-W(T))^{2}}{2\Sigma^{2}(T)}+\Gamma(T)\exp (-w)t\biggr )
\biggr ] dw, \]
where
\[ E(T,t)=\frac{\sigma(t)}{\epsilon_{0}} \]
is the current Young modulus.

We begin with matching experimental data for a polyurethane resin
($T_{\rm g}=80$~$^{\circ}$C).
A description of specimens and the experimental procedure can be found in
\cite{LG88}.
First, we fit observations in relaxation tests at various temperatures $T$.
Given a parameter $E_{0}$, the quantities $W$, $\Sigma$ and $\Gamma$
are found by using the steepest-descent algorithm.
Afterwards, the initial Young modulus $E_{0}$ is determined by
the least-squares technique.

Figure~1 demonstrates excellent agreement between experimental data and
results of numerical simulation.
The parameters $W$, $\Sigma$, $E_{0}$ and $\Gamma$ are plotted versus
the degree of supercooling $\Delta T=T_{\rm g}-T$ is Figures~2 and 3.
These figures show that far below the glass transition temperature,
the quantities $W$, $\Sigma$, $E_{0}$ and $\Gamma$
are fairly well approximated by the ``linear'' functions
\begin{equation}
W=a_{0}+a_{1}\Delta T,
\qquad
\Sigma=b_{0}+b_{1}\Delta T,
\qquad
E_{0}=c_{0}+c_{1}\Delta T,
\qquad
\log\Gamma =d_{0}+d_{1}\Delta T
\end{equation}
with adjustable parameters $a_{i}$, $b_{i}$, $c_{i}$ and $d_{i}$.
The values of these quantities at $T=75$~$^{\circ}$C, i.e. in the
close vicinity of the glass transition temperature, deviate from
dependences (22).
This may be explained by the fact that at a very small degree of
supercooling ($\Delta T=5$~K), the relaxation test was carried out
at the strain ($\epsilon=0.01$) that exceeded the yield strain.
As a result, experimental data (curve~4 in Figure~1) demonstrate
a combined effect of viscoelastic and viscoplastic deformations.
This issue has been previously discussed in \cite{QPG95}.

The parameters $W$ and $\Sigma$ increase with a decrease in temperature,
which is in good agreement with an assumption about an increase
in the ruggedness of the energy landscape with the growth of the degree
of supercooling \cite{AMF99}.
The initial Young modulus $E_{0}$ decreases with temperature
in accord with conventional observations for the effect of temperature
on elastic moduli.
Surprisingly, the relaxation rate $\Gamma$ decreases with temperature.
This conclusion does not, however, contradict the theory of thermally
activated processes, because the relaxation rate $\Gamma$ is defined
[see Eq. (3)] as the product of the attempt rate $\Gamma_{0}$
(which increases with temperature) by the
factor $\exp\Bigl [-\Omega(T)\Bigr ]$ that characterizes
the ascent of the reference energy level at temperature $T$
with respect to that at the glass transition temperature.

We now approximate experimental data in a tensile test with a constant
strain rate $\dot{\epsilon}_{0}$,
\begin{equation}
\epsilon(t)=\left \{\begin{array}{ll}
0, & t<0,\\
\dot{\epsilon}_{0}t, & t\geq 0.
\end{array}\right .
\end{equation}
For this purpose, we integrate Eqs. (14), (15), (19) and (20) with
the material parameters $W$, $\Sigma$ and $\Gamma$
determined in relaxation tests and find the parameters $g_{0}$
and $K_{i}$ which minimize deviations between results of numerical
simulation and experimental data.
Because the initial Young modulus $E_{0}$ found in relaxation tests
underestimates stresses at the initial region of the stress--strain
curve, the value of $E_{0}$ was found by fitting experimental data
in the interval $\epsilon\in [0,0.02]$.

First, we approximate the stress--strain curve at room
temperature ($T=25$~$^{\circ}$C) obtained at the strain rate
$\dot{\epsilon}_{0}=0.01$ min$^{-1}$.
The quantities $K_{i}$ and $g_{0}$ are found by applying
the steepest-descent procedure.
Afterwards, we fix the parameters $K_{i}$ and match experimental data
at other temperatures and other strain rates with the only
adjustable parameter $g_{0}$.
Figures~4 and 5 demonstrate good agreement between experimental data
and results of numerical analysis.
The initial threshold stress $g_{0}$ is plotted versus the strain rate
$\dot{\epsilon}_{0}$ in Figure~6.
This figure reveals that the dependence $g_{0}(\dot{\epsilon})$ is fairly
well approximated by the function (conventionally used
to describe the effect of strain rates on the yield stress)
\begin{equation}
g_{0}=\beta_{0}+\beta_{1}\log\dot{\epsilon}
\end{equation}
with adjustable parameters $\beta_{i}$.

We now repeat fitting experimental data for poly(methyl methacrylate)
($T_{\rm g}=105$~$^{\circ}$C) at room temperature.
Because of the lack of observations in relaxation tests and in tests with
constant strain rates for the same specimens,
we use experimental data in relaxation tests exposed in \cite{CFH81}
(the relaxation master curve corresponding to various aging times after
quench from 165~$^{\circ}$C)
and experimental data in tests with constant strain rates
measured in \cite{BBG99}.
We begin with matching observations in tensile relaxation test (21)
with the strain $\epsilon=0.005$.
Figure~7 demonstrates fair agreement between experimental data
and results of numerical simulation.
Using material parameters found in the relaxation test, we approximate
observations in tests with constant strain rates.
By analogy with the analysis of experimental data for polyurethane,
we determine the coefficients $K_{i}$ by fitting data in a test
with the strain rate $\dot{\epsilon}_{0}=0.01$ s$^{-1}$,
fix their values, and match observations for other strain rates
by using the only adjustable parameter $g_{0}$.
Figure~8 shows excellent agreement between experimental data
and results of numerical analysis.
The dependence $g_{0}(\dot{\epsilon})$ is depicted in Figure~6,
which reveals the high level of accuracy of Eq. (24).

To assess the effect of temperature on the viscoelastoplastic response
of poly(methyl metha\-cry\-late), we fit observations in relaxation
tests (data are adopted from \cite{MFR96})
and in tests with a constant strain rate (measurements are taken
from \cite{CTW98})
at various temperatures in the sub--$T_{\rm g}$ region.
Figure~9 reveals good agreement between experimental data and results of
numerical simulation.
Because experimental data are available in a relatively small interval
of time (3 decades), we reduce the number of adjustable parameters
by setting $W=0$ (which implies that only three constants are employed
to match observations).
The functions $\Sigma(T)$ and $\Gamma(T)$ are plotted in Figure~10,
which demonstrates that the parameter $\Sigma$ increases with
the degree of supercooling (in agreement with results presented in
Figure~2).
The rate of relaxation $\Gamma$ decreases with $\Delta T$, in agreement
with the theory of thermally activated processes.
It is worth noting that the parameter $\Sigma$ vanishes (the energy
landscape becomes homogeneous) at the critical temperature
$T_{\rm cr}=T_{\rm g}+4.7$~K, which is quite close to the critical
temperatures for other amorphous polymers, see \cite{Dro00}.
The initial Young modulus $E_{0}$ is plotted versus temperature $T$ in
Figure~11.
The dependence $E_{0}(\Delta T)$ is correctly described by Eq. (22),
where the coefficient $c_{i}$ take different values in the close
vicinity of $T_{\rm g}$, $\Delta T<20$~K, and far below the glass
transition point.
Observations in tests with constant strain rates are depicted
in Figure~12 together with results of numerical simulation
(the initial Young modulus $E_{0}(T)$ was taken directly from
the data in relaxation tests).
This figure also reveals fair agreement between experimental data and
the model's predictions.
Material parameters $g_{0}$ and $K_{i}$ for polyurethane and poly(methyl
methacrylate) are listed in Table~1 which shows that these quantities
have the same order of magnitude for the two polymers.

\section{Concluding remarks}

Constitutive equations have been derived for uniaxial isothermal
response of glassy polymers at small strains.
An amorphous polymer is treated as an ensemble of CRRs trapped
in their cages.
At low stress levels, relaxing regions are connected one with another,
which ensures that macro-strain in a specimen coincides with the
average micro-strain in a CRR.
When the average stress exceeds some threshold level, links between CRRs
break, and relaxing regions can slide one with respect to another
(which induced viscoplastic deformation at the micro-level).
Simple kinetic equations are proposed for changes in the threshold
stress and plastic strain.
The stress--strain relations are verified using experimental data
for a polyurethane resin and poly(methyl methacrylate) at various
temperatures and strain rates.
Fair agreement is established between experimental data and results
of numerical simulation.

\subsubsection*{Acknowledgement}
\noindent
This work was supported by the Israeli Ministry of Science
through grant 1202--1--98.

\newpage

\section*{Figure legends}

\begin{enumerate}
\item
The Young modulus $E$ GPa versus time $t$ s
for polyurethane at temperature $T$~$^{\circ}$C.
Circles: experimental data \cite{LG88}.
Solid lines: results of numerical simulation.
Curve~1: $T=25.0$;
curve~2: $T=35.0$;
curve~3: $T=50.0$;
curve~4: $T=75.0$

\item
The dimensionless parameters $W$ (unfilled circles)
and $\Sigma$ (filled circles) versus the degree of supercooling
$\Delta T$~$^{\circ}$C for polyurethane.
Circles: treatment of observations \cite{LG88}.
Solid lines: approximation of the experimental data by Eq. (22).
Curve~1: $a_{0}=-12.8947$, $a_{1}=0.7053$;
curve~2: $b_{0}=-9.4737$, $b_{1}=0.4263$

\item
The initial Young modulus $E_{0}$ GPa (unfilled circles)
and the relaxation rate $\Gamma$ s$^{-1}$ (filled circles) versus
the degree of supercooling $\Delta T$~$^{\circ}$C for polyurethane.
Circles: treatment of observations \cite{LG88}.
Solid lines: approximation of the experimental data by Eq. (22).
Curve~1: $c_{0}=1.8221$, $c_{1}=0.0100$;
curve~2: $d_{0}=-1.9310$, $d_{1}=0.0571$

\item
Stress $\sigma$ MPa versus strain $\epsilon$
for polyurethane loaded at temperature $T$~$^{\circ}$C
with the strain rate $\dot{\epsilon}_{0}=0.01$ min$^{-1}$.
Circles: experimental data \cite{LG88}.
Solid lines: results of numerical simulation.
Curve~1: $T=25.0$;
curve~2: $T=50.0$

\item
Stress $\sigma$ MPa versus strain $\epsilon$
for polyurethane loaded at $T=25$~$^{\circ}$C with the strain rate
$\dot{\epsilon}_{0}$ min$^{-1}$.
Circles: experimental data \cite{LG88}.
Solid lines: results of numerical simulation.
Curve~1: $\dot{\epsilon}_{0}=0.1$;
curve~2: $\dot{\epsilon}_{0}=0.01$;
curve~3: $\dot{\epsilon}_{0}=0.005$

\item
The initial threshold stress $g_{0}$
versus the rate of strain $\dot{\epsilon}_{0}$ min$^{-1}$
at $T=25$~$^{\circ}$C.
Circles: treatment of observations for a polyurethane resin
\cite{LG88} and poly(methyl methacrylate) \cite{BBG99}.
Solid line: approximation of the experimental data by Eq. (24).
Curve~1: PU, $a_{0}=88.096$ and $a_{1}=13.448$;
curve~2: PMMA, $a_{0}=97.922$ and $a_{1}=20.176$

\item
The Young modulus $E$ GPa versus time $t$ s
for poly(methyl methacrylate) at $T=25$~$^{\circ}$C.
Circles: experimental data \cite{CFH81}.
Solid line: results of numerical simulation with
$W_{0}=12.4$, $\Sigma_{0}=14.8$, $E_{0}=2.0442$ GPa
and $\Gamma=0.05$ s$^{-1}$

\item
Stress $\sigma$ MPa versus strain $\epsilon$
for poly(methyl methacrylate) loaded at $T=25$~$^{\circ}$C
with the strain rate $\dot{\epsilon}_{0}$ s$^{-1}$.
Circles: experimental data \cite{BBG99}.
Solid lines: results of numerical simulation.
Curve~1: $\dot{\epsilon}_{0}=10^{-2}$;
curve~2: $\dot{\epsilon}_{0}=5\times 10^{-3}$;
curve~3: $\dot{\epsilon}_{0}=10^{-3}$;
curve~4: $\dot{\epsilon}_{0}=5\times 10^{-4}$;
curve~5: $\dot{\epsilon}_{0}=10^{-4}$

\item
The Young modulus $E$ GPa versus time $t$ s
for poly(meth\-yl methacrylate) at temperature $T$~$^{\circ}$C.
Circles: experimental data \cite{MFR96}.
Solid lines: predictions of the model.
Curve~1: $T=50$;
curve~2: $T=75$;
curve~3: $T=85$;
curve~4: $T=90$;
curve~5: $T=100$

\item
The dimensionless parameter $\Sigma$
(unfilled circles) and the rate of relaxation $\Gamma$ s$^{-1}$
(filled circles) versus the degree of supercooling
$T$~$^{\circ}$C for poly(methyl methacrylate).
Circles: treatment of observations \cite{MFR96}.
Solid lines: approximation of the experimental data by Eq. (22).
Curve~1: $b_{0}=3.4931$, $b_{1}=0.7483$;
curve~2: $d_{0}=1.0853$, $d_{1}=-0.0564$

\item
The initial Young modulus $E_{0}$ GPa
versus the degree of supercooling $\Delta T$~$^{\circ}$C
for poly(meth\-yl methacrylate).
Circles: treatment of observations \cite{MFR96}.
Solid lines: approximation of the experimental data by Eq. (22).
Curve~1: $c_{0}=0.3217$, $c_{1}=0.0504$;
curve~2: $c_{0}=1.1763$, $c_{1}=0.0069$

\item
Stress $\sigma$ MPa versus strain $\epsilon$
for poly(methyl methacrylate) loaded at temperature $T$~$^{\circ}$C
with the strain rate $\dot{\epsilon}_{0}=0.1$ min$^{-1}$.
Circles: experimental data \cite{CTW98}.
Solid lines: results of numerical simulation.
Curve~1: $T=50$;
curve~2: $T=75$
\end{enumerate}

\newpage
\begin{table}[t]
\begin{center}
\caption{Adjustable parameters $g_{0}$ and $K_{i}$
for polyurethane and poly(methyl methacrylate)}
\vspace*{5 mm}


\end{center}
\vspace*{10 mm}

\caption{}
\end{figure}

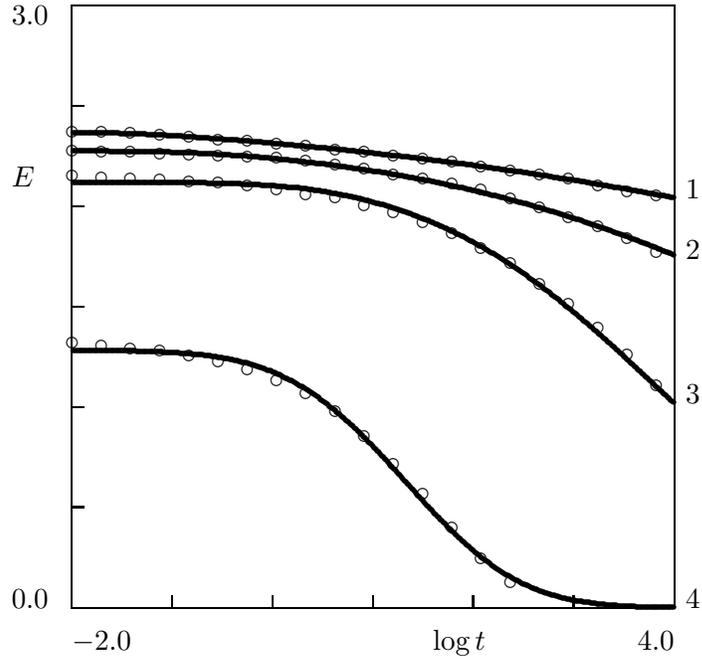
\begin{figure}[t]
\begin{center}
%
\end{center}
\vspace*{10 mm}

\caption{}
\end{figure}

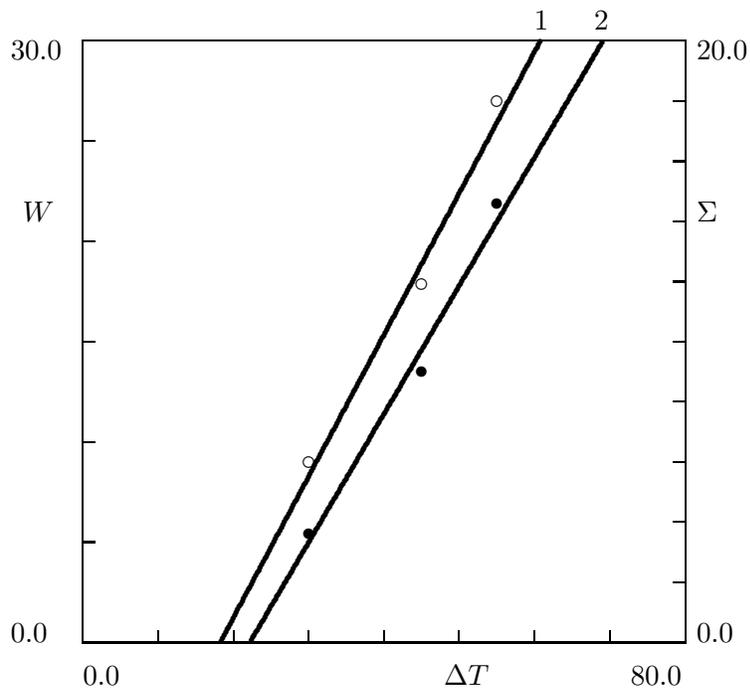
\begin{figure}[t]
\begin{center}
%
\end{center}
\vspace*{10 mm}

\caption{}
\end{figure}

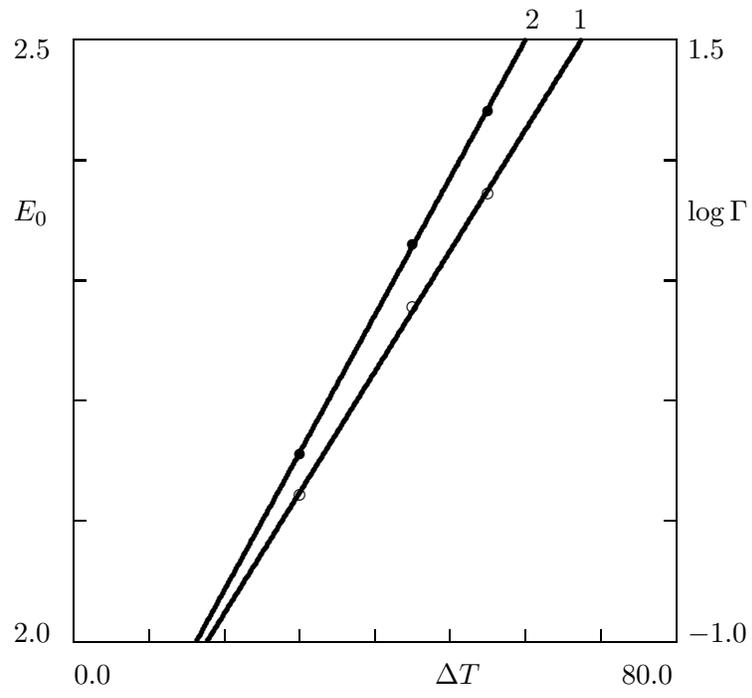
\begin{figure}[t]
\begin{center}
%
\end{center}
\vspace*{10 mm}

\caption{}
\end{figure}

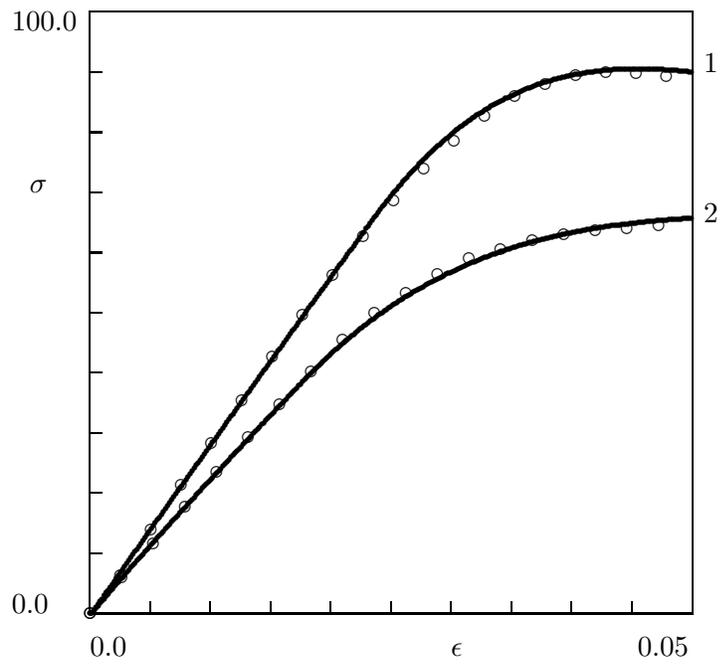
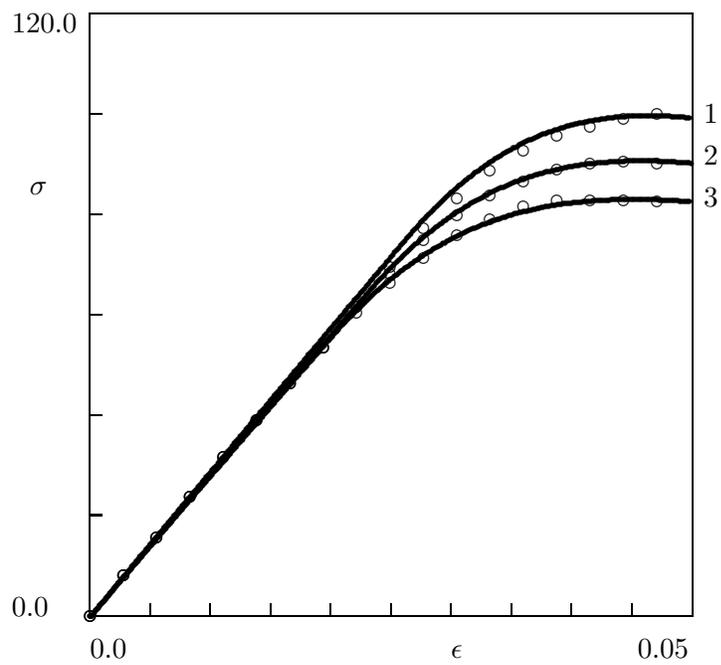
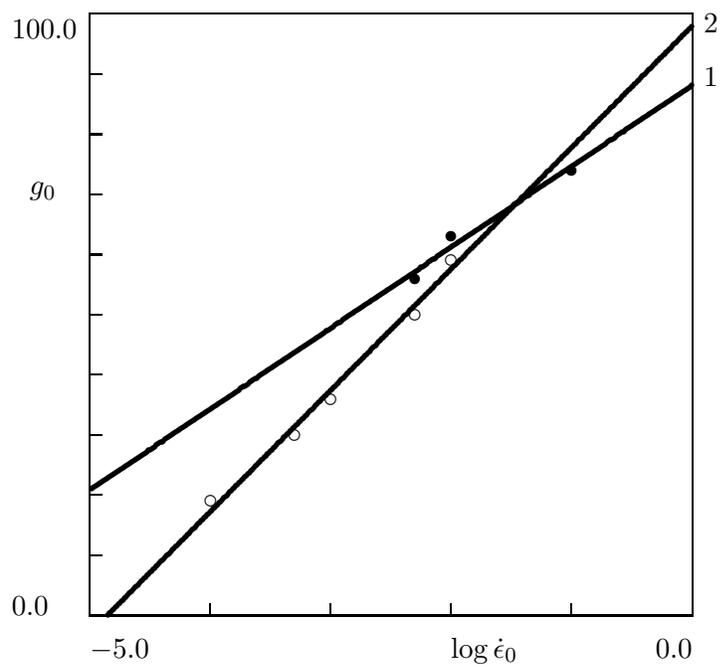
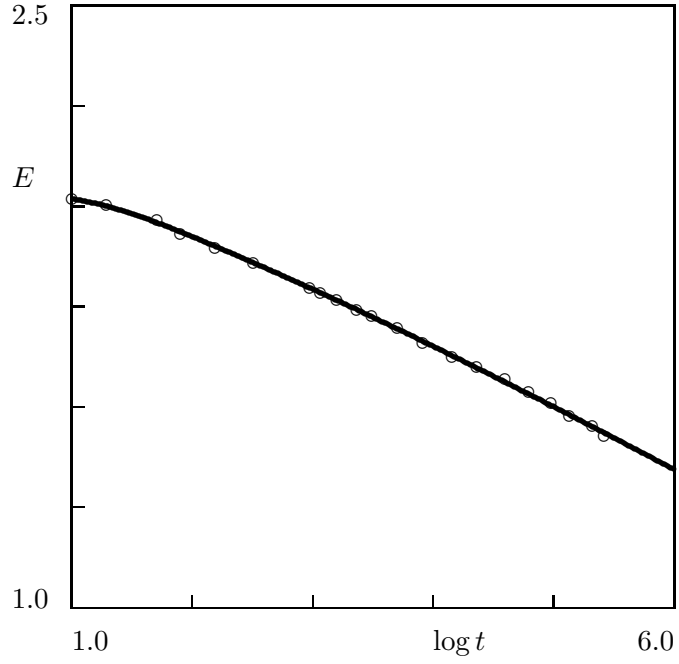
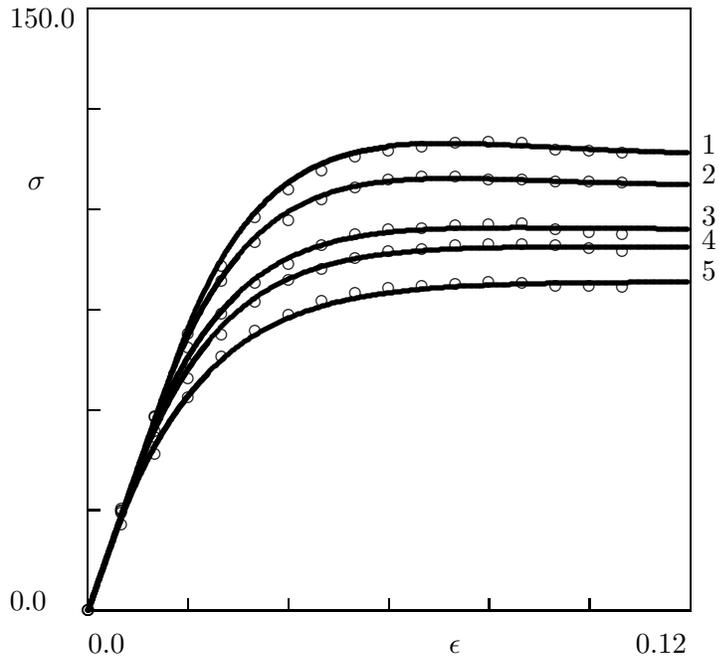
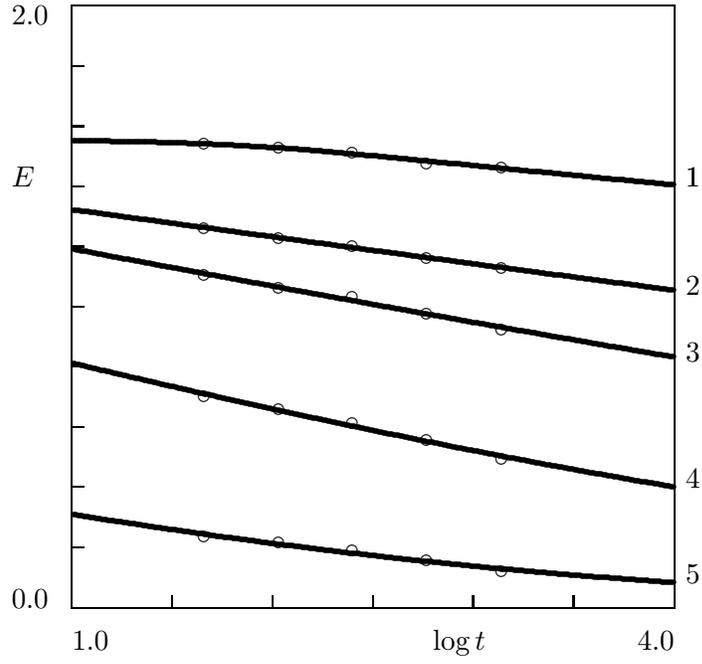
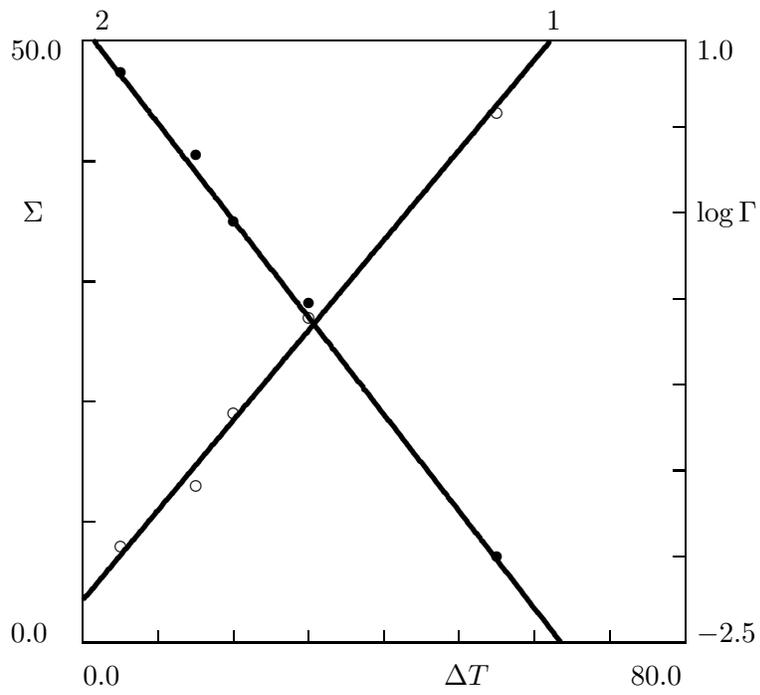
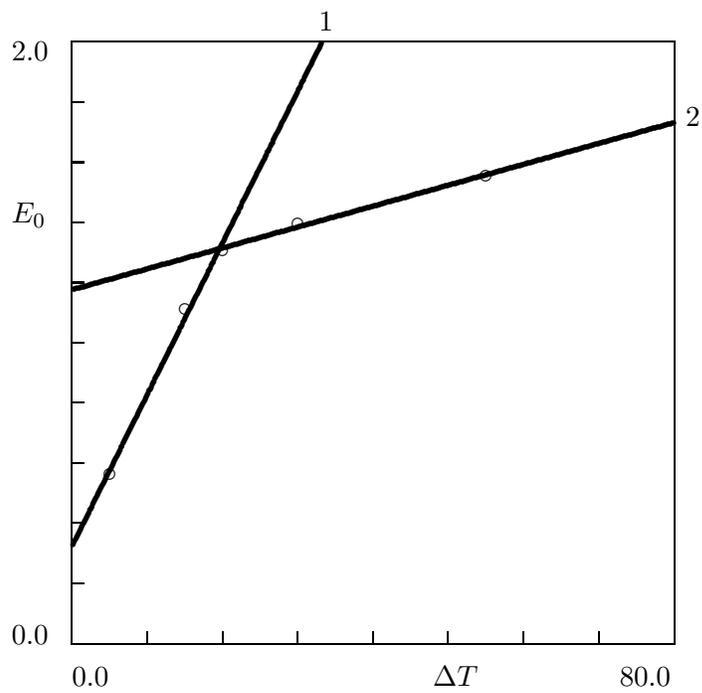
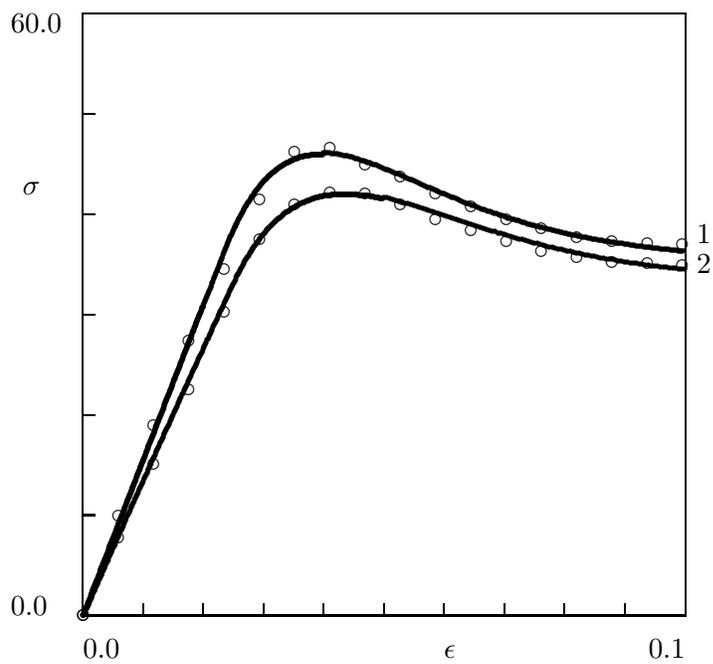
\end{document}